\documentclass[final,5p,times,twocolumn]{elsarticle}

\usepackage{amsfonts}
\usepackage{amssymb}
\usepackage{amsmath}
\usepackage[]{algpseudocode}
\usepackage{color}
\usepackage{algorithm}
\usepackage{graphicx}
\usepackage{pdfsync}
\usepackage[pdftex]{hyperref}
\usepackage{listings}

\newcommand{\drvec}{{\rm d}\vec{r}}

\definecolor{gray}{rgb}{0.4,0.4,0.4}
\definecolor{darkblue}{rgb}{0.0,0.0,0.6}
\definecolor{cyan}{rgb}{0.0,0.6,0.6}

\usepackage[usenames,dvipsnames,svgnames,table]{xcolor}

\lstdefinelanguage{XML}
{
	morestring=[b]",
	morestring=[s]{>}{<},
	morecomment=[s]{<?}{?>},
	stringstyle=\color{black},
	identifierstyle=\color{darkblue},
	keywordstyle=\color{cyan},
	morekeywords={xmlns,version,type}
}

\begin{document}

\title{Simflowny 3: An upgraded platform for scientific modelling and simulation}

\author[iac3,uib]{C.~Palenzuela\corref{cor1}}          \ead{carlos.palenzuela@uib.es}
\author[iac3]{B.~Mi\~{n}ano}              \ead{borja.minano@uib.es}
\author[iac3]{A.~Arbona}				  \ead{antoni.arbona@uib.es}
\author[iac3,uib]{C.~Bona-Casas}          \ead{carles.bona@uib.es}
\author[iac3,uib]{C.~Bona}                \ead{cbona@uib.es}
\author[iac3,uib]{J.~Mass\'o}             \ead{joan.masso@uib.es}
\cortext[cor1]{Corresponding author}
\address[iac3]{IAC$\,^3$, University of the Balearic Islands, Mateu Orfila, Cra. de Valldemossa km 7.5, 07122, Palma, Spain}
\address[uib]{Departament  de  F\'{\i}sica, University of the Balearic Islands and Institut  d'Estudis
Espacials  de  Catalunya,  Palma  de  Mallorca,  Baleares  E-07122,  Spain}

\begin{abstract}
Simflowny is an open platform which automatically generates efficient parallel code of scientific dynamical models for different simulation frameworks.  Here we present major  upgrades  on  this  software  to  support simultaneously a quite generic family of partial differential equations. These equations can be discretized using: (i) standard finite-difference for systems with derivatives up to any order, (ii) High-Resolution-Shock-Capturing methods to deal with shocks and discontinuities of balance law equations, and (iii) particle-based methods. We have improved the adaptive-mesh-refinement algorithms to preserve the convergence order of the numerical methods, which is a requirement for improving scalability. Finally, we have also extended our graphical user interface (GUI) to accommodate these and future  families  of  equations.  This  paper  summarizes  the  formal representation and implementation of these new families, providing several validation results.
\end{abstract}

\maketitle

{\bf PROGRAM SUMMARY}
  
\begin{small}
\noindent
{\em Program Title: }Simflowny                                 \\
{\em Licensing provisions: } Apache License, 2.0                      \\
{\em Programming language: } Java, C++ and JavaScript   \\
{\em Journal Reference of previous version:} Comput. Phys. Comm. 184 (2013) 2321--2331, Comput. Phys. Comm. 229 (2018), 170--181 \\
{\em Does the new version supersede the previous version?:} Yes  \\
{\em Reasons for the new version:} Additional features \\
{\em Summary of revisions:}\\
Expanded support for Partial Differential Equations, meshless particles 
and advanced Adaptive Mesh Refinement.\\
{\em Computer: }\\
  Simflowny runs in any computer with Docker \cite{dockerintro}. Installation details can be checked in the documentation of Simflowny \cite{simflownydoc}. It can also be compiled from scratch on any Linux system, provided dependences are properly installed as indicated in the documentation.\\
  The generated code runs on any Linux platform ranging from personal workstations to clusters and parallel supercomputers.  \\
{\em Nature of problem:}\\
Simflowny generates numerical simulation code for a wide range of models. \\
{\em Solution method:}\\
  Any discretization scheme based on either Finite Volume Methods, Finite Difference Methods, or meshless methods for Partial Differential Equations.\\
{\em Additional comments:}\\
  The software architecture is easily extensible for future additional model families and simulation frameworks.\\
Full documentation is available in the wiki home of the Simflowny project \cite{simflownydoc}. \\

\end{small}

\section{Introduction}

We present a significantly upgraded version 3 of Simflowny~\cite{Arbona20132321,ARBONA2018170},
an open platform for scientific dynamical models, composed of a Domain Specific Language (DSL) and a web based Integrated Development Environment (IDE) with a user friendly Graphical User Interface (GUI), which automatically generates efficient parallel code for simulation frameworks. Simflowny has a simple yet ambitious goal: a complete split of the physics (by introducing the concept of models and their associated problems) from the numerical 
methods necessary for a simulation (the discretization schemes), and the automatic generation of the final simulation code (where the parallel features and other capabilities of the chosen framework will be leveraged).

Historically, version 1 of Simflowny\cite{Arbona20132321} introduced the basic DSL, together with a family of PDEs and a first version of the GUI, and was able to automatically generate code for Cactus and SAMRAI. Version 2 extended the families of PDEs supported and introduced a completely revamped GUI, as well as support for Agent Based Models\cite{ARBONA2018170} with BOOST as the framework. In both papers, the DSL for different families of equations and IDE/GUI side of Simflowny were considered in depth, but the scalability of the final code has been specifically shown in the area of Numerical Relativity\cite{Palenzuela:2018sly}. This is an important issue, as Simflowny relies on the infrastructure of the framework for the efficiency of the generated parallel code and its scalability. This stronglys depend on the set of physical equations that one tries to simulate, and the results for the Einstein Equations are excellent.

This last version of Simflowny generates code for the SAMRAI infrastructure\cite{samrai,samrai_2016}. This does not mean that Simflowny can not generate code for other platforms (such as Cactus or BOOST, as in previous versions of Simflowny), but that SAMRAI is the only framework capable of supporting all the capabilities of Simflowny 3, as it provides efficient parallelization and adaptive mesh refinement (AMR), reaching exascale at least for some specific problems. Simflowny 3 is open source, and it is available in the form of compilable
source code and also as a Docker \cite{docker} container (see Simflowny's wiki for further details
\footnote{https://bitbucket.org/iac3/simflowny/wiki/Home}).

The main purpose of version 3 is to allow for the combination, into a single problem,  of all the different PDE discretization strategies available in previous versions, including: (i) standard finite-difference schemes, (ii) High-Resolution-Shock-Capturing(HRSC) methods for balance law systems, and (iii) particle methods to solve either a discrete set of particles or field evolution equations in an unstructured mesh represented by the particles. The combination of these discretization policies is performed both in the abstract sense (i.e., without yet referring to a specific software platform) and also practically by extending the SAMRAI simulation platform to support all these discretization methods.
Notice that an optimal platform to evolve a mesh-free system of particles might not be the most efficient way to solve PDEs on a structured grid, and vice versa. In particular, SAMRAI was not originally built to support
by default particle simulations. However, as one of our goals is to simultaneously evolve particles and fields with exascale-level parallelization and advanced AMR capabilities, we need a single platform supporting them both. 

Up to the best of our knowledge, we are not aware of any project that combines an IDE (with a GUI) and a DSL with automatic code generation that supports a quite generic family of PDE equations, particle simulations and AMR capabilities. This is a stringent combination of requirements but it is in no way complete, as Simflowny does not yet support the usage of Finite Element (FE) discretizations. We are aware of excellent DSL-based alternative projects which partially satisfy some of these requirements. For FE automatic computing and particles, one can use the combination of FEniCS\cite{fenics}, which uses a DSL based on a Unified Form language\cite{fenics_ufl} for weak formulations of PDEs\cite{Logg2007a}, with LEoPart\cite{leopart}, an add-on for FEniCS to integrate Lagrangian particle functionality with (Eulerian) mesh-based FE approaches. For finite differencing (FD), there is Devito\cite{devito-api,devito-compiler}, a DSL and code generation framework for the design of highly optimized FD kernels for use in inversion methods, which utilizes SymPy to allow the definition of operators from high-level symbolic equations. A combination of all these features would be a good basis for supporting wider user needs and we are currently considering these projects as potential candidates for supported frameworks in future versions of Simflowny.

The paper is organized as follows. The current status of Simflowny 3 and the pre-existent functionalities are summarized in section~\ref{section::status}. The new streamlined PDE architecture is described in Section~\ref{section::evolution}. Section~\ref{section::discretization} describes the support for discretization schemes in Simflowny 3, including particles, while Section ~\ref{section::amr} details the advanced AMR capabilities implemented through the SAMRAI infrastructure.
We illustrate these concepts in Section~\ref{section::examples} with problems based on a dam break, described with particle methods, and the evolution of charged particles in an self-consistent electromagnetic field, a scenario that Simflowny 2 could not have addressed.
We finish with some final remarks and conclusions in Section~\ref{section::conclusion}.

\section{Current status and preexistent functionality}
\label{section::status}

This section summarizes the functionalities of previous versions of Simflowny, which have been presented in detail in Refs.~\cite{Arbona20132321,ARBONA2018170}, together with a full description of the current status of the platform.

The DSL in Simflowny is based on an XML Schema Definition (XSD) representation. The XSD schemas
prescribe the structure of the XML documents for models, problems, and discretization schemes. Whenever algorithms need to be included in either of these XML documents, this is achieved through a specific markup language developed for Simflowny, called SimML (Simulation Markup Language). SimML includes a
full-blown rule specification language (technically speaking,
the language is Turing-complete, meaning that any possible
algorithm can be created with it). To insert mathematical expressions in the algorithms and documents, Simflowny leverages MathML, the standard markup language  for representing mathematical expressions.
 
In version 3, the DSL has been streamlined and 
currently supports a generalized evolution PDE paradigm, which 
includes any sort of spatial derivatives but only first order 
derivatives in time. As subcases of this generalized paradigm
we may single out: 
\begin{itemize}
    \item Partial Differential Equations (PDE) written in balance law form. This family was the only one supported in version 1 of Simflowny.    
    In this scenario the PDEs are written as an 
    evolution system containing only first order derivatives both in time and space. Space derivatives are always represented as the derivative of some flux function of the fields, allowing to use numerical schemes based on Finite Volume Methods to deal with shocks and discontinuities. The most typical example of balance law systems is the Euler equations, but wave equations, Maxwell equations or even Einstein equations can also be written in this form.
   \item Evolution PDEs, allowing for
    almost arbitrary forms of an evolution system by including spatial derivatives of any order, which are discretized with standard finite-difference operators. 
    The only restriction is that
    the system must be still first order in time. This is not really a strong restriction, as any PDE system can be reduced to a first-order one by introducing additional variables for the time derivatives of the evolved fields. Therefore, the restriction is just that the first-order reduction must be used, rather than the original higher-order version.    
    This family of equations was available in version 2 of Simflowny. One could write in this form some of the examples from balance law form (e.g., the Einstein, Maxwell and wave equations) but directly as a second order system in space. Additionally, it also includes many other equations like the heat equation, the Navier-Stokes-Korteweg equation (third order in space), the Cahn-Hilliard equation (fourth order in space) and the Phase-Field-Crystal equation (sixth order in space).
\end{itemize}

Simflowny 3 merges these two families into a single one, allowing any possible combination in the PDEs, in addition to the preexisting forms. Moreover, it allows for a representation of these PDEs either on a structured regular mesh, or on a set of (unstructured) moving point-particles. Obviously, spatial discretization methods, either finite difference or particle methods, will depend on the type of representation chosen by the user.

From the computer science point of view, 
Simflowny is built on the well-known three-tier architecture: (i) a presentation tier, implemented as a web browser based 
IDE with a GUI,
(ii) a logic tier, based on an application server, and (iii) a data tier, combining native XML databases with bulk data storage.
The current procedure to generate code is similar to 
the previous version. The process to convert a mathematical model 
into a numerical code can be split in four stages: 
\begin{enumerate}
\item  The representation of the mathematical model, which  contains  the PDE equations to be evolved.
\item The representation of the problem, which includes the mathematical model, the domain of the simulation, the analysis quantities and the initial and boundary conditions to be applied to evolution fields (in PDEs),
as well as the finalization condition.
\item The representation of the discrete scheme, which converts the continuous problem into a discrete one by defining the space and time discretization operators.
\item The generation of the code for the discrete problem
into the simulation framework. These frameworks will essentially play the role of a mesh/particle manager by setting the domain, distributing the usage of memory of the fields and parallelizing the workload among the different processors.  
\end{enumerate}
Notice that Simflowny's four stage structure allows us to achieve our stated main goal: a complete split of the physics from the numerical methods and from the parallelization/distribution issues, which are hidden in the framework. Therefore, we can use the same representation of the discrete problem (which involves the first three stages) in different simulation frameworks, which might allow for extended features, higher scalability and efficiency.

As in previous versions, Simflowny 3 provides support for the SAMRAI mesh
management toolkit~\cite{samrai,samrai_2016}.
Through SAMRAI, Simflowny 3 supports mesh-based discretization schemes
for PDEs, specifically Finite Difference Methods (FDM) and Finite Volume Methods (FVM). It also supports particle-based mesh-free Lagrangian methods (i.e., Smooth Particle Hydrodynamics, SPH).
Simflowny 3 takes full advantage of the parallelization capabilities of SAMRAI, and therefore any model developed on Simflowny can generate optimized
parallel code for this framework. For efficiency reasons, code responsible for communication is only generated
in the parts of the model which are not local: fluxes and spatial derivatives, while keeping source terms local.
Simflowny is thus abstracted from the implementation details that enable parallelization. Therefore, at code generation time the only input needed is where, when and what kind of data synchronization should be called. Parallelization in SAMRAI is provided through MPI using a domain decomposition paradigm; according to Flynn`s taxonomy the simulation codes are SPMD (Single Program, Multiple Data): all threads execute the same simulation code in a specific domain chunk. The number of threads in which the simulation is divided is set by users at runtime.
Furthermore, the capabilities of Adaptive Mesh Refinement are available in Simflowny by leveraging its implementation in SAMRAI.

\section{Evolution PDE}
\label{section::evolution}

In Simflowny 2 we had two separate families, such that a PDE could be written either in balance law form to be discretized with HRSC schemes, or as an Evolution PDE, with spatial derivatives of any order, to be discretized with standard finite-difference operators. In Simflowny 3 these two families are merged into a single Evolution PDE which covers all the possible combinations of the two options, both at the continuum and at the discretized level. Simflowny 3 is backwards compatible, in the sense that it is still possible to write any balance law equation as in version 2, but now it adds the possibility to combine features of balance law equations with the more flexible conditions of the former Evolution PDE family. Furthermore, Simflowny 3 also allows for fields to be represented by a set of particles, in addition (or combined) with the fields represented within a mesh from version 2.

Currently, the Evolution PDE family allows for arbitrary forms of evolution equations, including spatial derivatives of any order. The only restriction is that the system can only involve first time derivatives of the fields. This condition will allow us to use the Method of Lines (MoL), that provides a clean separation of the spatial and temporal discretization and help to ensure stability of the discretized system of equations.
In Simflowny 3, in the Evolution PDE, each field has an associated representation, namely either on a regular grid or by a mesh-free set of particles. Each Evolution PDE must be defined within a reference frame, which can be either Eulerian (i.e., fields evolve on a fixed mesh) or Lagrangian (i.e., fields co-move with the particle trajectories). Notice that fields on a Lagrangian frame can only be discretized by a set of particles, while Eulerian ones can indistinctly use a regular grid or a set of particles.

The fields represented on a regular grid follows an  evolution equation of the form:
\begin{equation*} 
\partial_t \bf{u} + \partial_i F^i (u) = S(u) + \cal{L (\bf{u},\partial_i \bf{u})} 
\end{equation*} 
where $\bf{u}$ is an array with all the evolved fields, $\bf{F}^i$ is the flux in the direction $x_i$, $\bf{S}$ is the source term and $\cal{L (\bf{u},\partial_i \bf{u})}$ is any operator depending on the fields and its spatial derivatives of any order. This form is certainly redundant, since the derivatives of the fluxes can always be expressed using the $\mathcal{L}$ operator. 
However, when fluxes are explicitly identified, the model allows for the application of specific discretization HRSC schemes to the fluxes to deal with possible discontinuities and shocks appearing in the solution.

The fields represented by a set of mesh-free particles, in addition to the previous PDE, will include evolution equations to describe the trajectories of such particles, namely
\begin{equation*} 
   \frac{d {\bf x}_p}{dt} = {\bf v}_p  ~~~~,~~~~
   \frac{d {\bf v}_p}{dt} = {\cal{R}}_p
\end{equation*} 
where $(\bf{x}_p, \bf{v}_p)$ are the position and  velocities of the particle $p$, and ${\cal{R}}_p$ 
is a free-function to define some specific trajectories.

As in the version 2 of Simflowny, operators with arbitrary derivatives of any order can be constructed by using recursive rules, which can be written formally as
\begin{equation*} {\cal L}_i^{(0)} = f_i(\bf{u}) \end{equation*} 
\begin{equation*} {\cal L}_m^{(n)} = \sum_{n=0}^{n-1} \sum_i g^i({\bf u}) \prod_j \partial_k \left( {\cal L}_l^{(n-1)} C^{ijkl}_{m} \right) \end{equation*} 
where \(f_i,g^i\) are arbitrary functions depending on \(\bf{u}\) but not on its derivatives, and \(C^{ijkl}_{m}\) is a generic matrix which in practice will have only one non-trivial component.
Notice that the new Evolution PDE family can also be expressed in a logical abstract language as
\begin{eqnarray} 
\partial_t \bf{u} + \partial_i F^i (u) &=& S(u) + \sum( f ({\bf u}) {\rm D}) \\
{\rm D} &=& \prod\partial_i(\rm g(\bf{u}) {\rm D}) \end{eqnarray}
where, similar to the former definition, f and g are arbitrary algebraic functions on \(\bf{u}\), and \({\rm D}\) is recursive term allowing a complex set of differential expressions. Both algebraic and recursive terms are optional at every level of the recursion.
Let us stress that it is possible to apply different discretization algorithms to different terms on these equations. For instance, one could define advective terms, which must be discretized with one-side finite-difference schemes. Consequently, the evolution equation terms are grouped by the so called Operators, which are just a tag/name for a group of terms which are going to be discretized in the same way.

\section{Discretization strategies}
\label{section::discretization}

One of the new features in Simflowny 3 is the support for particles, such that a set of PDEs can be discretized and simulated using mesh-free particle methods, even in combination with standard structured grid methods within the same simulation. Notice that the discretization of fields on a regular grid is given just by a formula depending on a fixed number of static neighboring points. However, the discretization of fields represented by a set of particles involves a dynamical set of moving neighbors that might change with time. 

\subsection{Spatial discretization}

Structured grid methods rely on the strong assumption that the fields evolve on a regular static grid which  does not change with time. Therefore, discrete spatial derivatives of a field $U$ can be written as a function depending on such field, evaluated at $s$ neighboring points, namely
\begin{eqnarray}\label{derU}
      (\partial_x U)_i = \sum_{j=-s}^{s} b_{i+j} U_{i+j}
\end{eqnarray}
where $b_{i+j}$ are constant coefficients and $i$ spans the regular grid points $x=i \Delta x$.

In balance law systems, the derivative of the fluxes can be treated by using a conservative discretization, namely
\begin{eqnarray}
  (\partial_x F)_i = \frac{1}{\Delta x} (F_{i+1/2} - F_{i-1/2}) ~~.
\end{eqnarray}
The reconstructed fluxes at the interfaces, $F_{i\pm1/2}$, need to be calculated with a monotonic reconstruction in order to deal with shocks and discontinuities, which can be written generically as
\begin{eqnarray}
   F_{i\pm1/2} = \sum_{j=-s}^{s} c_{i+j} F_{i+j}
\end{eqnarray}
where $c_{i+j}$ are again constant coefficients and $s$ the stencil of the method.

Particle methods are more generic and reduce to grid-methods when the set of particles is chosen to be regular and static. Here we will focus on short-range particle methods, where the interaction is mainly dominated by the close-by particles. A particle may interact with others in a relatively small finite range (i.e., its neighborhood), defined by an $n-$sphere surrounding the origin particle with a given radius. Following a kernel distribution, nearest particles have the strongest influence. It is still possible to write the discrete spatial derivatives, evaluated at the particle $a$, as a sum over the neighboring points, namely
\begin{eqnarray}
	(\partial_x U)_a = \sum_{b} \Delta V_b\, U_{b} \, W_{ab}({\bf r})
\end{eqnarray}
where $\Delta V_b$
is the volume element associated to the particle $b$
and $W_{ab}({\bf r})$ is the kernel function, depending 
on the distance between particles ${\bf r} = |{\bf x}_b - {\bf x}_a|$. The sum over $b$ spans all the particles in the radius of influence of the particle $a$, which are defined as those located at a distance ${\bf r}$ smaller than the interaction distance (see Appendix A).  

Depending on the specific model to be discretized, in addition to these operators, there might be also interactions between grid-based and particle-based fields:
\begin{itemize}
	\item \textbf{Particle to mesh interaction}. Particle-fields influence the grid-fields. In this case, the required particle-fields are mapped into the mesh by using a kernel interpolation from particles in the kernel range of the grid point.
	Therefore, the interpolation formula for a mesh node $n$ is approximated by the following summation over the neighboring particles:
	\begin{eqnarray}\label{Volume}
	u_n &=&  \sum_p \Delta V_p\,u_p \, W(\vec{r}_n-\vec{r}_p, h)
	\end{eqnarray}
	where, being $\vec{r}_n$ position of mesh node, summation goes over all the particles $p$, which have a position $\vec{r}_p$ and an associated volume $\Delta V_{p}$

	\item \textbf{Mesh to particle interaction}. Particle-fields require information from the grid-fields. At every particle, a linear Lagrange polynomial interpolation is performed with the nearest surrounding grid nodes to reconstruct that field on the particle.
	\begin{eqnarray}\label{Volume}
		u_p &=& \sum_n u_n l_n(\vec{r}_n) \\
		l_n(\vec{r}_n) &=& \prod_{j=0, n \neq j}^{n} \frac{\vec{r}_p - \vec{r}_j}{\vec{r}_n - \vec{r}_j}
	\end{eqnarray}
	where $n$ is the set of nodes and respectively $\vec{r}_n$ and $\vec{r}_p$ position of a node and the particle.
	
\end{itemize}

\subsection{Time discretization}

We use the Method of Lines to discretize the evolution PDE of the fields represented either on a structured grid or by a set of mesh-free particles. Grid-based methods always assume an Eulerian frame, where
the grid is fixed and the fields evolution, represented with the fixed-position time derivative operator $\partial_t$, occurs on this static grid. Particle-based methods instead are naturally semi-Lagrangian, in the sense that time derivative operator $D_t$ follows the particles, although spatial derivatives are yet computed in the Eulerian frame. 

Therefore, it might be required to convert between the  Eulerian time derivative into a Lagrangian one by using the transformation rule $\partial_t \to D_t - v^k \partial_k$. Let us consider
a generic equation for the field $U$, namely
\begin{equation}
\partial_t U = {\cal L}(U,\partial U)
\end{equation}
where ${\cal L}$ is a generic right-hand-side operator which can involve fields and their derivatives. This equation, in the Lagrangian frame, translates into
\begin{equation}\label{time_lagrangian}
D_t U - v^k \partial_k U = {\cal L}(U,\partial U)
\end{equation}
Simflowny allows the setting of a reference frame (either Lagrangian or Eulerian) for each evolution equation in the  model. If a Eulerian reference frame is set in a particle discretized equation, the extra term $v^k \partial_k U$ is automatically added to the equation. For Lagrangian particle equations, that term in not required. Note that grid-based discretizations are only allowed for equations with an Eulerian framework. Furthermore, Simflowny allows to specify multiple particle species with different models and evolve them independently. 

\subsection{Implementation details}

The Method of Lines allows an splitting of the spatial and time discretizations. For this reason, Simflowny also separates the spatial and time discretization schemes in two different sets, represented by specific documents in the IDE, and called respectively Spatial Operator Discretization and PDE discretization Schema. While the former prescribe all the spatial discretizations, the latter defines the integration of these abstract spatial discretizations into the time discretization.

The user can construct the specific procedure for one of the three different available discretizations (i.e., standard finite-difference, HRSC methods for fluxes or particle methods) in the PDE discretization Scheme. Notice that the HRSC methods require the characteristic speed information and involve complex algorithms, not appropriate for arbitrary-order derivative terms.
The evolution scheme (i.e., MOL with a specific time integrator and the spatial discretizations) is chosen in the Discretization Policy, together with the arbitrary-order derivative operator.

A spatial operator discretization document is used in Simflowny to define how the PDE arbitrary-order spatial derivatives are going to be discretized from the continuum expression. New spatial operator discretization documents are added from the document manager (see Ref.~\cite{ARBONA2018170}). The main core of this document lays within the Discretizations element. The RHS of equations might include terms having arbitrary-order derivatives and conservative flux terms. For the first terms, each derivative can be discretized either directly or recursively. By directly, a second order derivative would need a second order discretization. In a recursive discretization, the second order derivative could be discretized applying twice a first order discrete operator. Obviously, the result may be different. It is a good practice to always define a first order discretization, even when higher order discretizations are also defined, in order to guarantee the discretization of any arbitrary order PDE.

The algorithm presented in section~\ref{AMRalgorith_gridbased} shows how equations (\ref{derU}-\ref{time_lagrangian}) fit into the simulation evolution. At the physics level, the first step requires to perform the interpolation of fields lying in particles and mesh whose values are needed outside their reference frame. For example, a field \textbf{$A$} defined on the mesh reference frame may be used in particle equations. Consequently, a \textbf{$A_p$} field is interpolated in every particle so it can be read when needed. Then, the next step is to perform the spatial discretization, which is called to calculate the rhs in the skeleton algorithm. Finally, the final step is to integrate in time. If the simulation contain particles, one must perform the appropriate particle movement. The other instructions in the algorithm are related to refinement capabilities, explained in the next section.

\section{Adaptive Mesh Refinement scheme}
\label{section::amr}

The AMR algorithm is constructed by using basic blocks (i.e., routines and functions) provided by SAMRAI, as we summarize below.

\subsection{AMR algorithm for grid-based field}
\label{AMRalgorith_gridbased}

There are two refinement tagging strategies provided by SAMRAI and integrated in Simflowny: Fixed Mesh Refinement (FMR), where the user specifies statically a set of boxes where the refinement is located, and Adaptive Mesh Refinement (AMR), where the user sets a criteria used to calculate dynamically the cells to be refined. 
Both types of tagging strategies can be combined in the same simulation. The tagging during a simulation with AMR might change with time, needing new refinement areas or disposing of older ones. This re-meshing procedure is periodically run and it is parameterized for every simulation. 

If a new refinement box is added when the simulation has already started, the solution in these new points is interpolated from the coarser level, a procedure usually known as \textit{prolongation}.
The complementary procedure, called \textit{restriction}, consists on transferring the information from the fine levels to the coarser ones. Notice that the restriction procedure is straightforward for grid-based methods, as the data can be copied directly in the overlapping points from different levels (i.e., from fine to coarse).

The algorithm skeleton is as follows:
\begin{algorithm}
\caption{Evolution algorithm skeleton}
\begin{algorithmic}[1]
	\State initialization
	\State refinement\_tagging
	\While{not simulation end}
	\ForAll{level}
	\ForAll{time integration step}
	\State meshToparticle\_interpolation [Eqs. 8 and 9]
	\State particleTomesh\_interpolation [Eq. 7]
	\State calculate\_rhs [Eqs. 3, 4, 5 and 6]
	\State integrate\_time [Eqs. 10 and 11]
	\State move\_particles
	\If{last time integration step}
	\State restrict from level to level-1
	\State synchronization(level - 1)
	\EndIf	   
	\State synchronization(level)
	\State prolong from level-1 to level (prepare ghost zones)
	\State calculate\_physical\_boundaries
	\EndFor
	\If{has to regrid(level)}
	\State refinement tagging(level)
	\EndIf
	\EndFor
	\EndWhile
\end{algorithmic}
\end{algorithm}

\begin{figure}[h]
	\includegraphics[width=4cm]{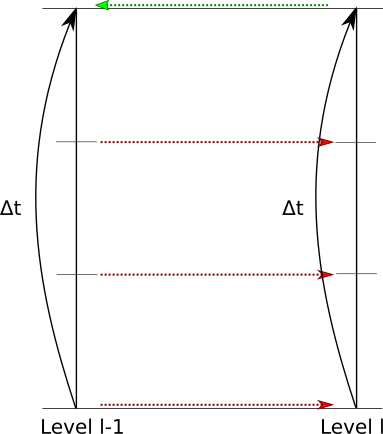}
	\includegraphics[width=4cm]{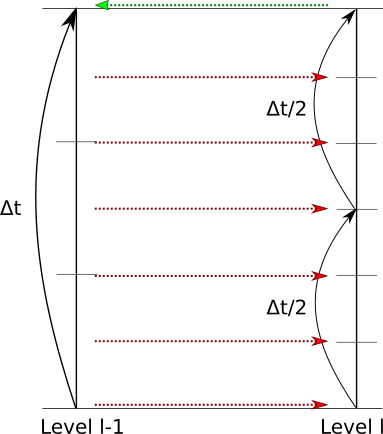}
	\caption{Synchronization between two levels with no sub-cycling (left) and with sub-cycling (right), using a Runge-Kutta with three sub-steps. Prolongation between levels is denoted with red arrows, while that restriction is denoted with green ones.}
	\label{figure:subcycling}
\end{figure}

The previous algorithm is similar for every level. However, prolongation, restriction and number of executions depend on whether sub-cycling in time is active. 
When using sub-cycling time integration, each level in the hierarchy runs for the longer possible $\Delta t$ while satisfying the Courant-Friedrich-Levy (CFL) condition to ensure stability. This implies that if the ratio between spatial resolutions on two consecutive refinement levels is $r$, the fine level will have a smaller timestep $\Delta t/r$, so the evolution on the children level must be repeated $r$-times to reach parent $\Delta t$. Sub-cycling requires a larger number of prolongation steps (i.e., red arrows) compared to the case without sub-cycling, as it is shown in Figure~\ref{figure:subcycling} for a third order Runge-Kutta (RK) time integrator. Due to the time misalignment on the RK sub-steps, a time interpolation must be performed in the coarser level, which might lead to a reduction of the accuracy if it is not performed carefully \cite{Mongwane:2015hja,Palenzuela:2018sly}. Although the prolongation is straightforward in the no sub-cycling case, the coarser levels run slower than they could, as the value of $\Delta t$ must fit to the Courant constraint in the finest level.
Currently, there are two available time integration strategies in Simflowny: with or without sub-cycling.

\subsection{AMR algorithm for particle-based field}

The concept of AMR for particles is more diffuse, as one could
add more particles from the beginning or the dynamics could focus them in a particular region, effectively increasing the accuracy of the solution. However, our aim 
is to add instantly more particles in a region where the grid has been refined. The challenges of running multi-resolution particle simulations in general imply processes of creation and destruction of particles. We will follow instead a simpler method commonly known as Multi Level Multi Domain(MLMD) approach~\cite{INNOCENTI:2013}.

The MLMD approach consists on a collection of different levels where particles are evolved independently by following the full physical description, and then interlocked to the others through an exchange of  information. Particles are created in each level of the grid they belong as it is created, and they remain bounded to this level from there on, not being allowed to perform a transition to other coarser or finer levels. Refined particles are destroyed when they exit their original domain, and new refined particles are created on the level from the coarse grid particle distribution at the boundary cells of the refined grids.

The algorithm of particle splitting creates, for each particle, a number of children particles given by a refinement ratio times the dimension. For instance, in a two-dimensional simulation with refinement ratio of $2$, each single coarse particle would split into $4$ fine particles. These new particles are shifted in space, keeping the same physical fields as parent particle $pp$. The volume of the parent particle is evenly distributed all over the new set of children particles $cp$.

\begin{eqnarray}\label{Volume}
V_{pp} &=& \sum_{cp} V_{cp}
\end{eqnarray}

More detail and properties of this method are explained in detail in ~\cite{LAPENTA:2002} (Algorithm S1).

An interesting characteristic of MLMD systems is that interaction between coarse and fine levels only occurs at fine boundary zones. Coarser particles create new finer particles at areas enforcing continuity, however, those finer particles (so as coarser particles) are free to evolve. Consequently, a different evolution across the levels is possible in the overlap areas simulated by different levels. 

The possibility of obtaining different dynamics across the levels, within the limits of boundary continuity, is considered a point of strength in MLMD, since it shows the capability of refined levels to evolve according to dynamics not accessible, because of the reduced resolution, to the coarser levels. This characteristic is one of the major differences from classic AMR systems: while AMR aims at consistency in the overlap area across the levels, MLMD only enforces particle continuity at grid interfaces and aims an efficient interlocking between the levels in the overlap areas. 

\section{Examples}
\label{section::examples}

The new functionalities on particle-based methods are shown by presenting two non-trivial examples. The first one is the damn break, which contains only fields described by particles. The second example is the consistent evolution of charge particles interacting with an electromagnetic field, involving both particles and grid-based fields, as well as the interaction between both of them. Notice that every document (i.e., schema, model and problem) used in the examples of this article is available in the Simflowny database\footnote{Default Simflowny installation provides a database with several models, including the ones described in this paper}.

\subsection{Dam break}

%
\begin{figure}[h]\centering
	\includegraphics[width=5cm]{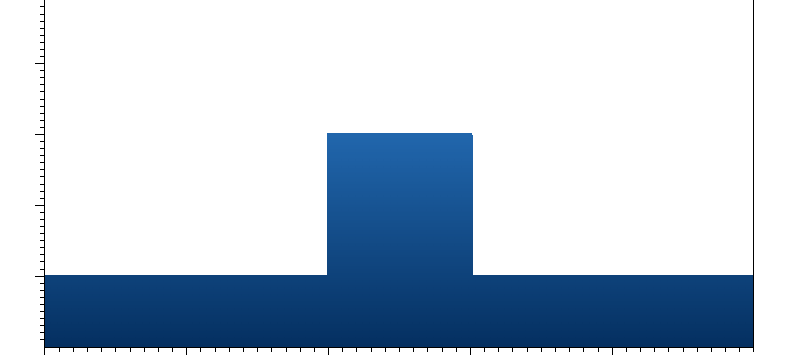}\\
	\includegraphics[width=5cm]{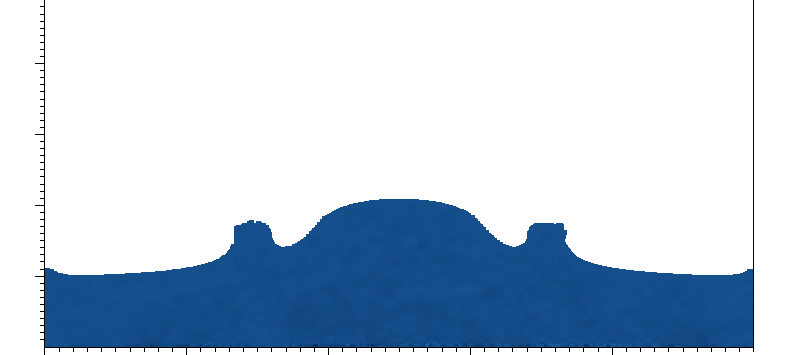}\\
	\includegraphics[width=5cm]{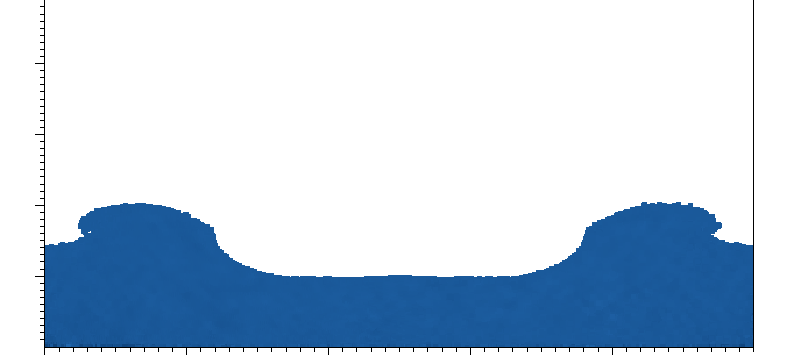}\\
	\includegraphics[width=5cm]{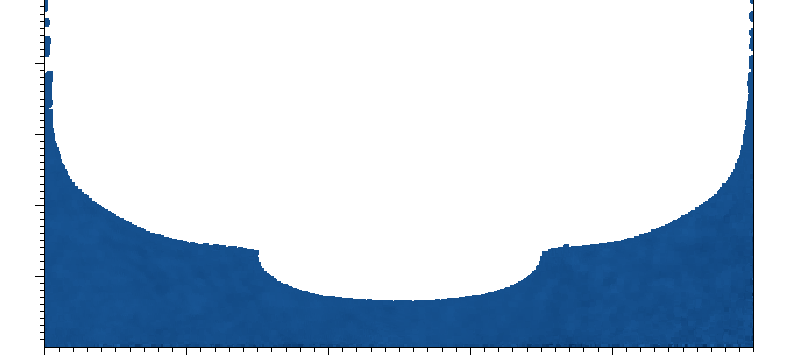}\\
	\includegraphics[width=5cm]{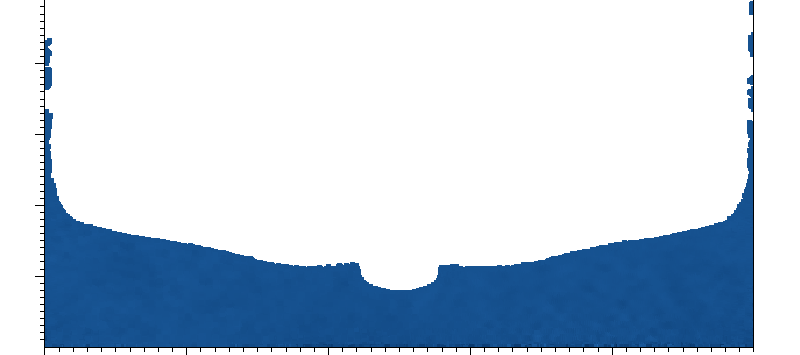}\\
	\includegraphics[width=5cm]{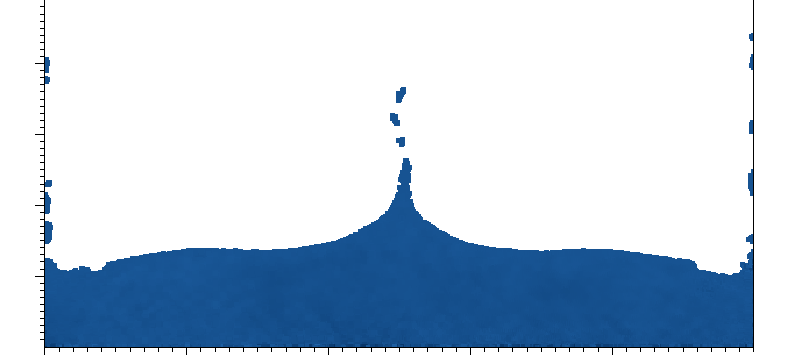}\\
	\includegraphics[width=5cm]{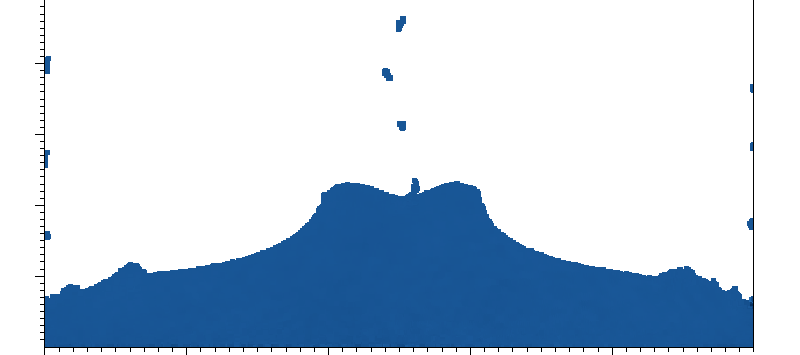}
	\caption{Simulation evolution for Dam Break with AMR with $\Delta t$ of $0.5$.}
	\label{figure:DamBreak}
\end{figure}

The first test is the dam break, a family of problems which consists on the study of a free falling mass of water with a given configuration, which starts at an initial location and flows all over the simulation domain, interacting with the environment. This problem is discretized with a particle-based method (Smoothed Particle Hydrodynamics, or SPH).
\begin{figure}[h]\centering
	\includegraphics[width=6cm]{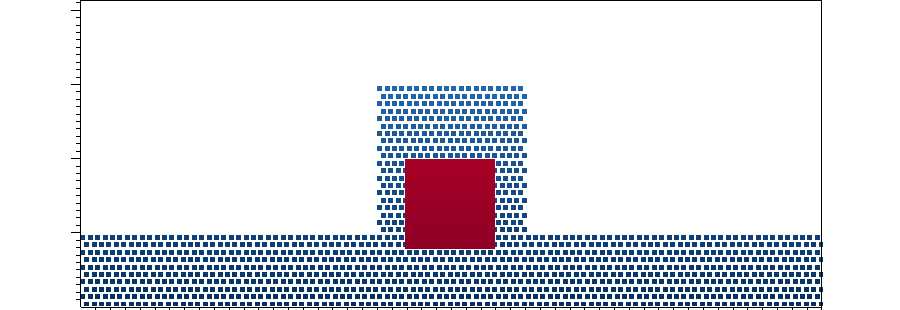}\\
	\includegraphics[width=6cm]{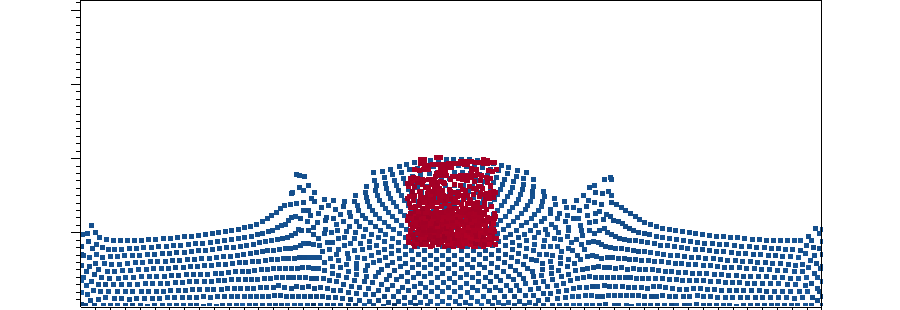}\\
	\includegraphics[width=6cm]{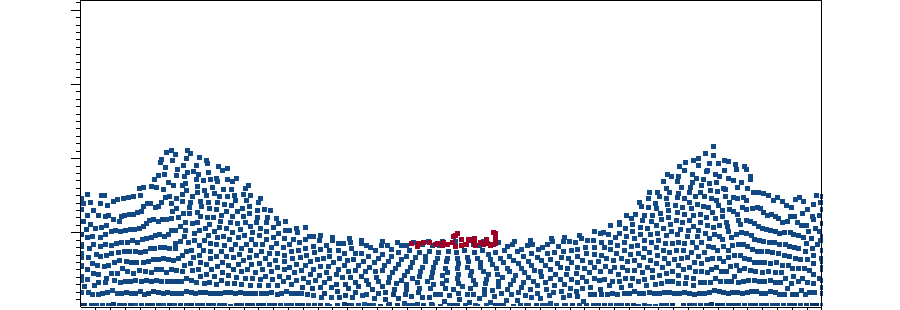}\\
	\includegraphics[width=6cm]{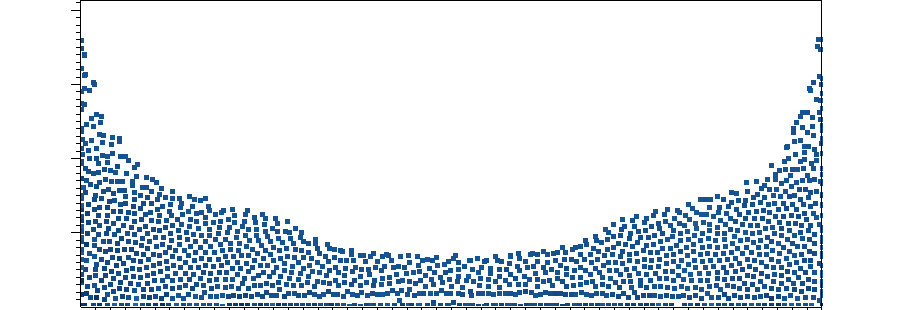}\\
	\includegraphics[width=6cm]{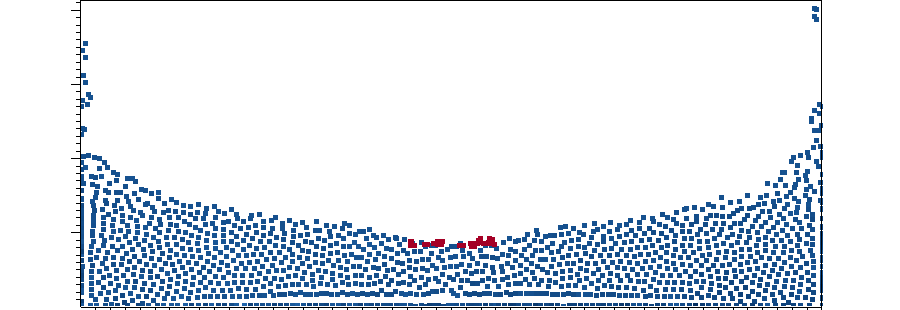}\\
	\includegraphics[width=6cm]{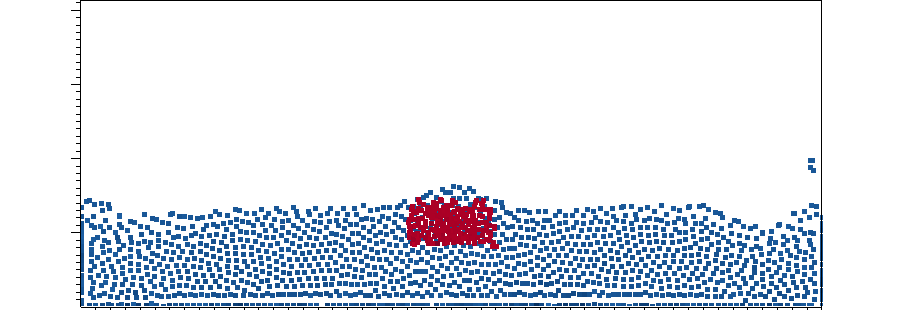}\\
	\includegraphics[width=6cm]{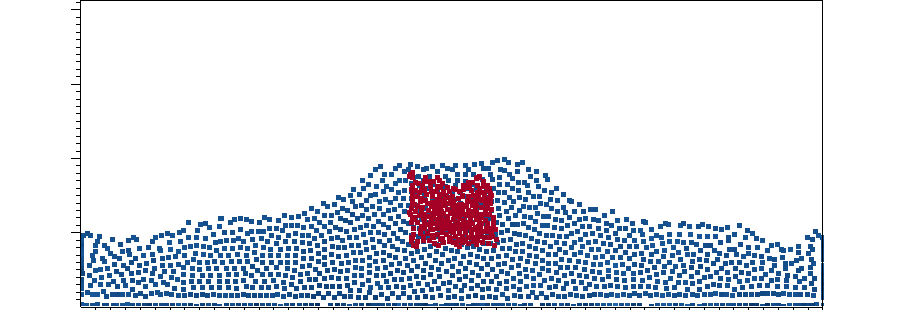}
	\caption{Simulation evolution for Dam Break with AMR with $\Delta t$ of $0.5$. Coarse particles in blue, fine particles in red.}
	\label{figure:DamBreakAMR}
\end{figure}
In this problem, water is described by the Navier-Stokes equations for an almost incompressible fluid (i.e., Stokes's stress constitutive equation), under the force of gravity, namely
\begin{eqnarray}
\frac{d \rho}{d t} &=& - \rho \nabla \cdot \vec{v} 
\\
\frac{d \vec{v}}{d t} &=& - \frac{1}{\rho} \nabla P 
+ \frac{\mu}{\rho} \nabla^2 \vec{v} + \vec{g}
\end{eqnarray}
where $\rho$ is the density, $\vec{v}$ the velocity vector, $\mu$ is the dynamic viscosity, $\vec{g}=(0,9.8)$ is the (constant) gravity vector, and the pressure $P$ is given by the equation of state $P=P_0 + (\frac{\rho}{\rho_0}^\gamma-1) \frac{\rho_0 c_0^2}{\gamma}$ with $\gamma=7$, $c_0=65$ and $\rho_0=1028$.

We consider a problem in a 2D domain in Cartesian coordinates $(x,y)$, with a initial setting consisting on a bed of water 10 units long and 1 tall. On the top center of that bed there is a square of water with side 2 (see top panel of Figure~\ref{figure:DamBreak}). Initial values for density and velocity are obtained assuming an equilibrium configuration, namely
\begin{equation}
\rho (t=0) = \rho_0 \frac{1.0 + \rho_0 g_y (H - y) \gamma}{\rho_0 c_0^2}^\frac{1}{\gamma} ~~,~~
\vec{v} (t=0) = 0
\end{equation}
The boundary conditions in this simulation are simple reflective walls. There is no physical wall, but just a change of sign on the velocity component perpendicular to the wall when a particle crosses the boundary, so that the particle is repelled back into the domain.

The discretization is performed by using the Method of Lines, with a Predictor-Corrector scheme for the time integration. The spatial derivatives are discretized by using particle methods, in particular the Smoothed-Particle Hydrodynamics(SPH) approach~\cite{SPH_review}.
The total number of particles in this run is 21600.
The dynamics of the flow, with a time sequence from the initial break on, is displayed in Figure~\ref{figure:DamBreak}.

This problem is also convenient to show the behavior of particle multi-resolution. For this example, there are 1400 particles on the coarsest level. We place a fixed refined box between $x=[4.5,5.5]$ and $y=[0.8,2.0]$ in the previous domain, containing 144 coarse particles, which are split following the MLMD algorithm into 576 particles on the finest level. As it can be observed in Figure~\ref{figure:DamBreakAMR}, the initial refined particles start to fall until all of them disappear from the refined-box region. When the water flow comes again to the refined-box area, new fine particles are created and evolve freely inside that area.

This example can be reproduced easily in Simflowny by following the tutorial in https://bitbucket.org/iac3/simflowny/wiki/DamBreakTutorial.

\subsection{Particles in an electromagnetic field}

We consider now two different species of charged particles (electrons and ions) moving in a self-generated electromagnetic field (EM). This system can be modeled by using Maxwell equations with a current source produced by the movement of these charged particles, which move under the influence of a Lorentz force. The equations of motion of the system for multiple species (in our case $s=\{ions,electrons\}$) are given by
\begin{eqnarray}
\partial_t \vec{E} &=& c^2 \nabla \times \vec{B} - \vec{J}/\epsilon_0
\\
\partial_t \vec{B} &=& - \nabla \times \vec{E}
\\
\frac{d \rho_s}{d t} &=& - \rho_s \nabla \cdot \vec{v}_s 
\\
\frac{d \vec{v}_s}{d t} &=& (q_s/m_s) \left( \vec{E} + \vec{v}_s \times \vec{B} \right)
\\
\frac{d \vec{x}_s}{d t} &=& \vec{v}_s
\end{eqnarray}
where $\rho_s$ is the electric charge density of each species, 
$\vec{v}_s$
its velocity and $\vec{x}_s$ its position. Here $q_s/m_s$ is a parameter of the model, different for each species, that determines the ratio of charge and mass of the particles being considered.
The total charge density $\rho$ and the total
electric current density $\vec{J}$,  sources of Maxwell field equations, are computed from the particle distribution as
\begin{eqnarray}
\rho &=& \sum_{s} \rho_s = \rho_e + \rho_i
\\
\vec{J} &=& \sum_s \rho_s \vec{v}_s = \rho_e \vec{v}_e + \rho_i \vec{v}_i
\end{eqnarray}
Notice that the EM fields are coupled to the particles through the sources, while the particle trajectories depend directly on the EM fields. 

The EM fields are discretized on a three-dimensional grid, while  charged particles are naturally described through particle-fields. Notice that this model has therefore no  direct equivalent one including only grid-based fields.

\begin{figure}[h]
	\centering
	\includegraphics[width=8cm]{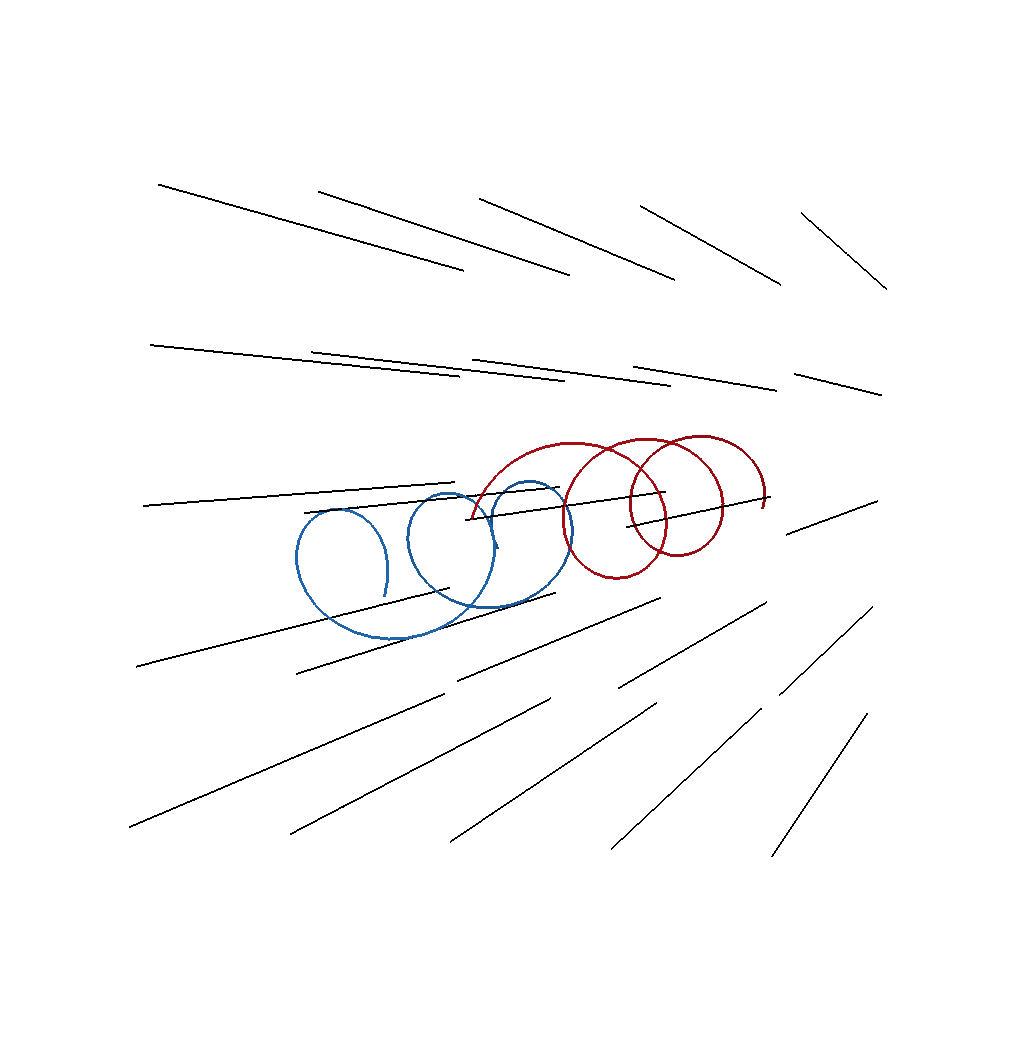}
	\caption{S5}
	\label{figure:pics2}
\end{figure}

For this test simulation we set 183 particles (108 electrons and 75 ions), alternating electrons with ions at the initial time. The initial conditions for fields are given by
\begin{eqnarray}
 \rho_e = - \rho_i = \rho_0 ~~&,&~~ 
\vec{v}_s = (0.1, 0.02, 0) \\
\vec{J}_s = \vec{E}_s = 0 ~~&,&~~
\vec{B}_s = (1, 0, 0 )  \nonumber
\end{eqnarray}

The discretization of this model also relies on the Method of Lines, using a fourth-order Runge-Kutta for the time integration. The spatial derivatives are discretized by using a Smoothed-Particle Hydrodynamics (SPH) scheme for the particle-based fields, while
the grid-based ones are discretized with High-Resolution-Shock-Capturing methods. We use the same specific method as in Ref.~\cite{2020arXiv200207554L}, namely a flux-splitting approach~\cite{shu98} combined with the MP5 reconstruction~\cite{suresh97}. 

For the sake of clarity, the trajectories of two charged particles, one of each species, are displayed in Figure~\ref{figure:pics2}. As expected, they follow an spiral trajectory (blue and red) around the magnetic field lines (in black).

\begin{figure}[h]\centering
	\includegraphics[width=8cm]{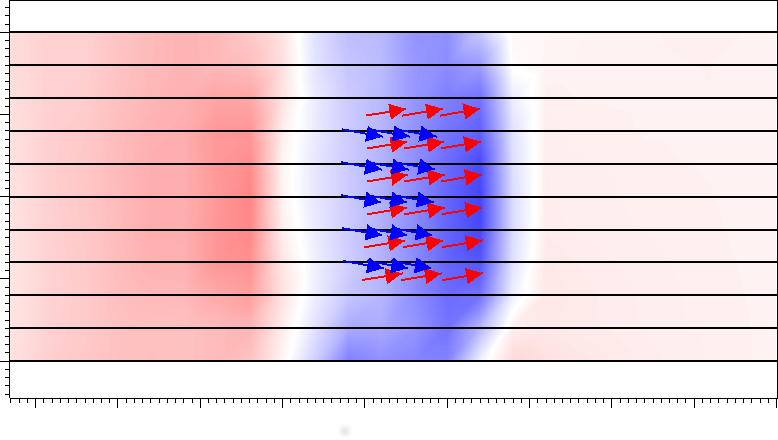}
	\caption{Lorentz force experiment. Ions (red) and electrons (blue) have influence on the Electric field (z component in picture). Horizontal black lines represent the magnetic field.}
	\label{figure:pics}
\end{figure}

It is easier to analyze the dynamics of the particles by displaying a two-dimensional slice. In Figure~\ref{figure:pics} the velocity vectors are shown, for each type of particle, together with the magnetic field lines and the z-component of the electric field. The initial set of particles move following a helical path, each species rotating in opposite sense as they move along the z-direction, and modifying the electric and magnetic fields as they move around.

\begin{figure}[h]\centering
	\includegraphics[width=8cm]{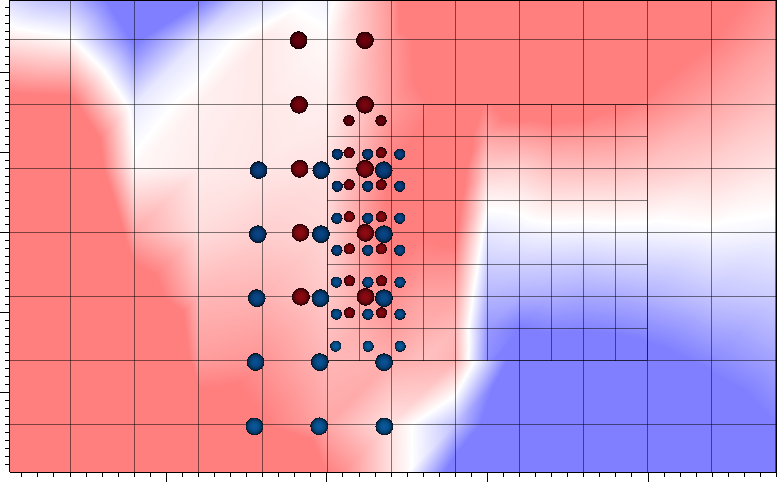}
	\caption{Lorentz force experiment with refinement. Ions (red) and Electrons (blue) have different sizes, smaller particles in finer level, larger in coarsest level. }
	\label{figure:pics_AMR}
\end{figure}

We can also test the multiple resolution particle features by setting a refined box in the center of our domain. In Figure~\ref{figure:pics_AMR} such case is displayed, where for visualization purposes the particles in the coarse grid have a size bigger than the (more numerous) particles defined in the finest level. Notice that particles split to create new ones as they enter into the refinement area.

As in the previous case, this example can be reproduced easily in Simflowny by following the tutorial in https://bitbucket.org/iac3/simflowny/wiki/PICTutorial

\section{Conclusions and future work}
\label{section::conclusion}

We have presented version 3 of Simflowny, an open platform which automatically is capable of generating efficient parallel code of scientific dynamical models for different simulation frameworks. 
We have introduced a new generic family of PDEs, together with several changes in the platform to discretize it with either standard finite differences, HRSC methods or mesh-less particle based methods. We have also implemented necessary improvements for advanced adaptive mesh refinement and several adaptations to the GUI. 
We have illustrated and validated the new generic family and the particle methods with two examples: the dam break and charged particles on an electromagnetic field. 

Simflowny 3 has now full support for a wide family of PDE equations and combines structured grids with mesh-less particles, all with advanced AMR, but it still lacks support for other extended methods like Finite Elements. 
A graphical IDE is an advantage for most users, but more advanced programmers might prefer simpler APIs not based on XML. For this reason, we are reevaluating our roadmap for future versions. On the one hand, we are following the advances in other DSL based projects for automatic generation of code\cite{fenics,leopart,devito-api}, as they can provide new frameworks for our platform, which might support also Finite Elements. Our semantic simulation language SimML, together with our GUI web-based IDE, already has all the underpinnings for producing code for these projects. On the other hand, we also plan to extend the support for other programmatical interfaces. Finally, we will also continue expanding the open database of available models.

\section*{Acknowledgements}
We acknowledge support from the Spanish Ministry of Economy and Competitiveness grant AYA2016-80289-P (AEI/FEDER, UE). 
We acknowledge PRACE for awarding us access to
MareNostrum at Barcelona Supercomputing Center (BSC), Spain, with the time granted through the $17^{th}$ PRACE regular call (project Tier-0 GEEFBNSM, P.I. CP).

\appendix

\section{Summary on particle discretization}

Our particle methods rely on Smooth-Particle-Hydrodynamics, where the particle discretization is an interpolation method that allow us to write down any field as a function of its values computed in any finite set of arbitrary points (i.e., the particles). Generically, any field $A(\vec{r})$ can be interpolated by using the integral formula
\begin{equation}\label{interp_integration}
<A(\vec{r})> = \int A(\vec{r}') \, W(\vec{r}-\vec{r}', h) \drvec' \; ,
\end{equation}
where the integration is performed in all the space coordinates and  $W(\vec{r}-\vec{r}', h)$ is the \emph{kernel} function, which depends on the smoothing length $h$ and must satisfy the following two basic conditions,
\begin{equation}
\int W(\vec{r}-\vec{r}', h) \drvec' = 1
~~~~~~,~~~~~~
\lim_{h\to 0} W(\vec{r}-\vec{r}', h) = \delta(\vec{r}-\vec{r}') \; .
\end{equation}

Therefore, the interpolation formula Eq.~(\ref{interp_integration}) is approximated by the following summation over the neighboring particles:
\begin{equation}
<A(\vec{r})> = \sum_b \Delta V_{b}\,A(\vec{r}_b) \, W(\vec{r}-\vec{r}_b, h)  ,
\end{equation}
where summation goes over all the particles $b$, which have a position $\vec{r}_b$, velocity $\vec{v}_b$ and an associated volume $\Delta V_{b}$.

One of the main advantages of particle methods is that we can build a differentiable interpolation of any field by using its values on the particles and a differentiable analytical \emph{kernel}. Consequently, derivatives of the field can be obtained without using finite differences on a grid. For instance, in order to compute $\nabla A$, we can just take the derivative of  Eq.~(\ref{interp_integration}), namely
\begin{equation} \label{eq:nablaA_orig}
<\nabla A(\vec{r})> = \sum_b \Delta V_{b}\, A(\vec{r}_b) \, \nabla W(\vec{r}-\vec{r}_b, h)  .
\end{equation}

The previous expressions can be used to compute the field and its gradient at the $a$, namely
\begin{eqnarray} \label{eq:nablaA_1}
<A_a> &=& \sum_b  \Delta V_{b}\, A_b \, W_{ab} , \\
<\nabla A_a> &=& \sum_b \Delta V_{b}\, A_b \nabla_a W_{ab}
\end{eqnarray}
where we have simplified the notation by defining $A_b \equiv A(\vec{r}_b)$ and $W_{ab} \equiv W(\vec{r}_a-\vec{r}_b, h)$.
If the \emph{kernel} is radially symmetric, then $W(\vec{r}_a-\vec{r}_b,h) = W(\vec{r}_b-\vec{r}_a,h) = W(|\vec{r}_b-\vec{r}_a|,h) $, that is, $W_{ab} = W_{ba}$.

One way to prevent inconsistencies near the boundaries, where there is not an isotropic distribution of particles, is by a suitable normalization of the previous particle formulas. For instance, the averaged value of the function can be normalized as:
\begin{equation}
<A_a> = \frac{\sum_{b}  \Delta V_b\, A_b W_{ab}}{\sum_b \Delta V_b\, W_{ab} } ~~.
\end{equation}
Notice that if the density $\rho$ is an evolved field, the associated volume can be estimated as $\Delta V_b=m_b/\rho_b$, where $m_b$ is the particle mass and $\rho_b$ its associated density.

The first derivative can be estimated in a similar way,
\begin{equation}
<\nabla A_a > = \frac{ \sum_{b} \Delta V_{b} \, \left[ A_b - A_a  \right] \nabla_a W_{ab} }{ \sum_{b} \Delta V_{b}\, \left[\vec{r}_b - \vec{r}_a \right] \nabla W_{ab} }
\end{equation}
which satisfies that it is zero for a constant field
and constant for a linear one.

\section{Details on particle implementation}

Currently, Simflowny generates code for SAMRAI, whose current implementation is exclusively based on structured meshes. This framework does not provide support for particles, but provides an open generic type to allow users to implement any desired structure in cells. 
Taking advantage of this functionality, it has been possible to implement particle support. The code generated by Simflowny extends the SAMRAI functionality as follows:

\begin{itemize}
	\item Every cell in the mesh contains an object with the list of particles located in that cell. There is one list for each particle species.
	\item Each particle species contains its respective information (fields, variables, and unique id) and methods for accessing that information.
	\item Methods have been developed to output particle information to disk in order to visualize simulation results.
	\item Methods for particle interpolation between levels have been added following the directives of MLMD, explained in Section~\ref{section::amr}.
\end{itemize}

Depending on its  position, each particle belongs to a certain single cell. Provided particle cell locality, getting the particle's neighborhood consists on looping over close cells and comparing distances with particles on those cells. The number of close particles to cover can be calculated from the kernel radius of particle-based scheme  and underlying mesh cell resolution.
Notice that these methods of binding lists of particles to each cell are also widely used in other simulation areas, such as molecular simulations\cite{Mattson}. 

The following text is an extract of the internal XML code internally representing the discretization policy from the example \textit{Particles in an electromagnetic field}. 

\lstset{language=XML}
\begin{lstlisting}
<?xml version="1.0" encoding="utf-8"?>
<mms:PDEDiscretizationPolicy ...>
   ...
   <mms:meshVariables>
      <mms:meshVariable>Bx</mms:meshVariable>
      <mms:meshVariable>By</mms:meshVariable>
      <mms:meshVariable>Bz</mms:meshVariable>
      <mms:meshVariable>Ex</mms:meshVariable>
      <mms:meshVariable>Ey</mms:meshVariable>
      <mms:meshVariable>Ez</mms:meshVariable>
   </mms:meshVariables>
   <mms:particleVariables>
      <mms:species>
      <mms:name>ion</mms:name>
      <mms:particleVariable>rho_i</mms:particleVariable>
      <mms:particleVariable>vx_i</mms:particleVariable>
      <mms:particleVariable>vy_i</mms:particleVariable>
      <mms:particleVariable>vz_i</mms:particleVariable>
      <mms:particleVariable>Jx_i</mms:particleVariable>
      <mms:particleVariable>Jy_i</mms:particleVariable>
      <mms:particleVariable>Jz_i</mms:particleVariable>
      <mms:velocities>
         <mms:velocity>
            <mms:component>x</mms:component>
            <mms:formula>
               <mt:math>
                  <mt:ci>vx_i</mt:ci>
               </mt:math>
            </mms:formula>
         </mms:velocity>
         ...
      </mms:velocities>
      <mms:volume>
         <mt:math>
            <mt:apply>
               <mt:divide/>
               <mt:ci>q</mt:ci>
               <mt:ci>rho_i</mt:ci>
            </mt:apply>
         </mt:math>
      </mms:volume>
   </mms:species>
   <mms:species>
      <mms:name>electron</mms:name>
      <mms:particleVariable>rho_e</mms:particleVariable>
      ...
      <mms:velocities>
         ...
      </mms:velocities>
      <mms:volume>
         ...
      </mms:volume>
   </mms:species>
   ...
\end{lstlisting}

The previous extract shows the electric and magnetic field components (Ex, Ey Ez, Bx, By, Bz) being defined as mesh variables; and then two particle species (ion and electron) defined, each one with their own fields, velocity components and volume. This XML description is internal, as Simflowny provides a GUI to ease data introduction. A complete User Guide can be found in our wiki page\footnote{https://bitbucket.org/iac3/simflowny/wiki/UserGuide}.

\end{document}